%% file: Main.tex
\documentclass[journal]{IEEEtran}

\usepackage{cite}
\usepackage{amsfonts,amsmath,amssymb}
\usepackage{graphicx,color,epsfig,rotating}
\usepackage{epstopdf} 
\usepackage{latexsym}
\usepackage{subfigure}
\usepackage{cite}
\usepackage{epic,eepic}
\usepackage{algorithm}
\usepackage{algorithmic}
\usepackage{float}
\usepackage{multirow}
\usepackage{graphicx,subfigure}
\def\BibTeX{{\rm B\kern-.05em{\sc i\kern-.025em b}\kern-.08em    T\kern-.1667em\lower.7ex\hbox{E}\kern-.125emX}}
\usepackage{bbm}
\usepackage{bm}
\usepackage[export]{adjustbox}

\allowdisplaybreaks

\usepackage{amsthm}
\theoremstyle{plain}

\input{macro_by_Justin}

\begin{document}
\title{\huge{Decentralized Covert Routing in Heterogeneous Networks Using Reinforcement Learning}}
\author{ \IEEEauthorblockN{Justin Kong,~\emph{Senior Member, IEEE}, Terrence J. Moore~\emph{Member, IEEE}, and \\ Fikadu T. Dagefu,~\emph{Senior Member, IEEE} } \\

\thanks{
Justin Kong, Fikadu T. Dagefu, and Terrence J. Moore are with the  U.S. Army Combat Capabilities Development Command (DEVCOM) Army Research Laboratory, Adelphi, MD 20783, USA (e-mail: justin.s.kong.civ@army.mil; fikadu.t.dagefu.civ@army.mil; terrence.j.moore.civ@army.mil)
}

}

\maketitle

\begin{abstract} 

This letter investigates covert routing communications in a heterogeneous network where a source transmits confidential data to a destination with the aid of relaying nodes where each transmitter judiciously chooses one modality among multiple communication modalities. We develop a novel reinforcement learning-based covert routing algorithm that finds a route from the source to the destination where each node identifies its next hop and modality only based on the local feedback information received from its neighboring nodes. We show based on numerical simulations that the proposed covert routing strategy has only negligible performance loss compared to the optimal centralized routing scheme.

\end{abstract}


\IEEEpeerreviewmaketitle

\section{Introduction}

\IEEEPARstart{C}{overt} communication that hides the very existence of wireless communication signals from potential adversaries has been extensively studied for security-sensitive networks. This is especially important when preventing adversaries from intercepting confidential data (i.e., the \textit{wiretap} channel problem~\cite{Wyner:75}) is not sufficient to ensure strong security~\cite{Hero:03}. 
The information theoretical covert communication analysis of an additive white Gaussian noise (AWGN) channel was examined in~\cite{Bash:13}.
The authors in~\cite{Kong:21} developed a covert communication technique for an intelligent reflecting surface (IRS)-assisted network by jointly optimizing the transmit probability, power at the transmitter and the excitation matrix at the IRS. 
Furthermore, channel inversion power control-based covert communications was explored in~\cite{Ma:21}.

When a direct link between a source and a destination cannot be established, relay nodes are often used to create a multi-hop link. 
For this case, considering the wiretap problem, the work in~\cite{Wang:18b} proposed a joint route and power allocation method that minimizes the communication outage probability under the constraint of a secrecy outage probability threshold. 
In~\cite{Wang:20}, for a network with multiple relaying nodes that are equidistantly placed on a line from a source to a destination, the optimization of the throughput of covert communication for protection against an adversarial node on an unmanned aerial vehicle was investigated.
In addition, for covert communication, joint path and power optimization for the throughput maximization and delay minimization based on a bound on the detection error probability (DEP) was devised in~\cite{Sheikholeslami:18}.

Q-learning, which is a type of reinforcement learning, has been widely adopted in wireless communications as a tool for solving various routing problems for complex wireless networks~\cite{Mammeri:19}.
Unlike the centralized schemes in~\cite{Wang:18b,Wang:20,Sheikholeslami:18} where a centralized node collects all information and makes a decision for all nodes, Q-learning based approaches enable decentralized routing algorithms where each node decides its own action based on local information.   
The study in~\cite{Jung:17} introduced a Q-learning-based geographic routing to enhance the network performance of unmanned robotic networks. 
A stochastic shortest-path problem for wireless sensor networks was tackled by utilizing Q-learning~\cite{Xia:19}.
In~\cite{Arafat:22}, a Q-learning-based topology-aware routing strategy with enhanced communication reliability for flying ad hoc networks was presented.

Heterogeneous networks (HetNets), where multiple independent networks are employed to improve throughput, security as well as coverage, have also been investigated~\cite{Wu:18b}.
The wiretap secrecy rate maximization for a two-tier downlink HetNet consisting of one macrocell and multiple femtocells was examined in~\cite{Lv:15}.
Also, the authors in~\cite{Xu:19} developed a security-aware energy-efficient resource allocation technique for multi-homing HetNets. 
For an air-to-ground network with omnidirectional microwave and directional millimeter wave communication modalities, numerical transmit power optimization and mode selection methods for covert communication were proposed in~\cite{Zhang:23}.
In addition, the joint optimal resource allocation and modality selection scheme for covert communication that maximizes the DEP while satisfying a requirement on the throughput at an intended receiver was devised for a HetNet with low-very high frequency and microwave frequency communication modalities in~\cite{Kong:22WCNC}.

It should be emphasized that the previous studies on secure wireless communication in~\cite{Wang:18b,Wang:20,Sheikholeslami:18,Lv:15,Xu:19,Zhang:23,Kong:22WCNC} only focused either on routing with a single communication modality or single-hop communication with multiple modalities. 
Hence, it is important to investigate covert routing techniques for HetNets.
In a centralized HetNet, a central node can optimize route and modality based on the global channel information that is received from all nodes~\cite{Kong:24}. 
However, the centralized scenario may not be practical as it requires high overhead which makes it more vulnerable from an adversarial detection point of view. For some applications, it is also desirable to avoid reliance on a single point of failure. Despite the need for decentralized covert routing algorithms for HetNets, such routing strategies have not been explored.

In this letter, we present a novel decentralized Q-learning-based covert (Q-covert) routing technique for HetNets that simultaneously identifies a route from a source to a destination and the optimal modality for each transmitting node in the route with the goal of maximizing the end-to-end DEP at an adversary (Willie) while satisfying a throughput requirement at the destination. 
Note that Q-learning has not been used for routing problems for covert communications. To the best of our knowledge, this is the first work that proposes Q-learning based covert routing.
The contributions of this letter are:
\begin{itemize}
\item We formulate the problem by defining the action space at each node for a joint next-hop and optimal modality selection, and specify the cost of each single-hop link for the end-to-end DEP maximization. 

\item We develop a novel Q-covert routing algorithm where a transmitting node chooses a next-hop and a communication modality for its transmission exploiting its estimate on the end-to-end DEP, which is obtained based on local feedback information received from neighboring nodes. 

\item We validate, through numerical simulations, that the proposed decentralized Q-covert routing scheme exhibits almost identical performance with a centralized optimal approach and significantly outperforms naive routing methods.

\end{itemize}

\section{Network model and problem formulation}

\subsection{Network Model}

We consider a HetNet consisting of a source, a destination, and an adversary Willie along with multiple other nodes from which relay nodes can be chosen to establish a route between the source and the destination. All nodes in the network are equipped with $M$ communication modalities where each modality has its own unique operating frequency and channel characteristics. 
Here, Willie has a radiometer that covers the frequency bands used by all $M$ modalities and tries to detect the existence of communication in the network.

Let us define $\Psi$ by the set of all possible routes from the source to the destination.
We denote $\psi = (h_1,\dots,h_{N_{\psi}}) \in \Psi$ as a route where $h_i=(T_{h_i},R_{h_i})$ stands for the single-hop link between transmitter~$T_{h_i}$ and receiver~$R_{h_i}$, and $N_{\psi}$ means the number of hops of the route $\psi$.
Here, the source and destination are $T_{h_1}$ and $R_{h_{N_{\psi}}}$, respectively.
In this HetNet, each of the transmitting nodes adaptively chooses one of the $M$ communication modalities based on the wireless environment and the communication requirement.

A communication slot for any transmitter/receiver pair single-hop communication is composed of a block of $L$ channel uses.
For any given slot, the transmitter decides whether to send data to the receiver or not with equal \textit{a priori} probability.
Then, for a single-hop link~$h$, when transmitter~$T_h$ selects modality~$m_h$ from multiple modalities for the transmission of the data symbol $x[l] \sim \CC(0,1)$ in the $l$-th channel use, the received signals at receiver~$R_h$ and Willie are respectively written as 
\begin{align} \label{eq:Received_Receiver}
	&y_{T_h,R_h}^{(m_h)}[l] = \sqrt{P} g_{T_h,R_h}^{(m_h)} x[l] + n_{T_h,R_h}^{(m_h)}[l], \\ \label{eq:Received_Willie}
	&y_{T_h,\text{W}}^{(m_h)}[l] = \sqrt{P} g_{T_h,\text{W}}^{(m_h)} x[l] + n_{T_h,\text{W}}^{(m_h)}[l],  
\end{align}
for $l=1,\dots,L$, where $P$ is the transmit power at all transmitters.
Here, $g_{T_h,R_h}^{(m_h)}$ and $g_{T_h,\text{W}}^{(m_h)}$, respectively, indicate for the channels from transmitter~$T_h$ to receiver~$R_h$ and to Willie for modality~$m_h$.
Also, $n_{T_h,R_h}^{(m_h)}[l]\sim\cC\cN(0,\Omega^{(m_h)}N_0)$ and $n_{T_h,\text{W}}^{(m_h)}[l]\sim\cC\cN(0,\Omega^{(m_h)}N_0)$ respectively specify the AWGN at receiver~$R_h$ and Willie for modality~$m_h$ where $\Omega^{(m_h)}$ and $N_0$ account for the bandwidth for modality $m_h$ and noise power spectral density, respectively.

\subsection{Detection at Willie}

We assume that Willie knows the optimized route from a source to a destination $\psi = (h_1,\dots,h_{N_\psi})$, utilized modalities along the routes $\{ m_h\}_{h\in\psi}$, channels from transmitters to Willie $\{g_{T_h,\text{W}}^{(m_h)}\}_{h\in\psi}$, transmit power $P$, and bandwidths $\{\Omega^{(m_h)}\}$. 
It should be noted that Willie may not have all of these information in practice. However, in this paper, we assume the worst case from a covertness point of view by providing Willie the capability to employ a highly sophisticated detection technique.

Willie attempts to detect the presence of the transmissions along the route $\psi$ by employing a hypothesis test on each of the single-hop communications constituting the route. For a single-hop link $h \in \psi$, Willie distinguishes the two hypotheses, the null hypothesis $\cH_h^{(0)}$ in which there is no transmission and the alternative hypothesis $\cH_h^{(1)}$ in which there exists a transmission. Using the average received signal strength $\bar{y}_{T_h,\text{W}}^{(m_h)} = \frac{1}{L} \sum_{l=1}^L \big| 	y_{T_h,\text{W}}^{(m_h)}[l] \big|^2$, Willie makes a binary decision about whether there is communication over link $h$ by performing a threshold test as follows~\cite{Bash:13,Kong:21,Ma:21}:
\begin{align} \label{eq:Willie_Decision}
	\bar{y}_{T_h,\text{W}}^{(m_h)} \underset{\cD_h^{(0)}}{\overset{\cD_h^{(1)}}{\gtrless}} \delta_h,
\end{align}
where $\delta_h$ stands for the detection threshold for link $h$, and $\cD_h^{(0)}$ and $\cD_h^{(1)}$ respectively denote the decisions in favor of $\cH_h^{(0)}$ and $\cH_h^{(1)}$.

As a transmitter decides whether to send data or not with equal \textit{a priori} probability, the DEP at Willie for a single-hop link $h$ is given by
\begin{align} \label{eq:DEP_def}
	&\tP_{\text{DEP},h} = \tP_{\text{MD},h} + \tP_{\text{FA},h},
\end{align}
where $\tP_{\text{MD},h} = \PP( \cD_h^{(0)} |  \cH_h^{(1)} )$ and $\tP_{\text{FA},h} = \PP( \cD_h^{(1)} |  \cH_h^{(0)} )$ are the missed detection probability and false alarm probability, respectively. 
Here, with the optimal detection threshold $\delta_h$, we can express the DEP in~\eqref{eq:DEP_def} as~\cite{Kong:22WCNC}
\begin{align} \label{eq:DEP_with_gammafuncs}
	\tP_{\text{DEP},h} = 1 - &\frac{1}{\Gamma(L)}  \Bigg(   \gamma\bigg( L, L \Big( 1 + X^{(m_h)} \Big) \ln \Big( 1 + \frac{1}{X^{(m_h)}} \Big) \bigg)   \nonumber \\
    & -   \gamma\bigg(   L, L  X^{(m_h)}    \ln \Big( 1 + \frac{1}{X^{(m_h)}} \Big)              \bigg)      \Bigg), 
\end{align}
where $\Gamma(s)=\int_0^\infty t^{s-1} e^{-t} \d t$ and $\gamma(s,x) = \int_0^x t^{s-1} e^{-t} \d t$ signify the gamma function and lower incomplete gamma function, respectively, and $X^{(m_h)} \triangleq \frac{\Omega^{(m_h)}N_0}{P | g_{T_h,\text{W}}^{(m_h)} |^2}$.

Willie will fail to detect the communication in a route $\psi$ if Willie makes wrong decisions for all hops in the route. Therefore, the end-to-end DEP of route $\psi$ is
\begin{align} \label{eq:DEP_route}
	\tP_{\text{DEP}}(\psi) = \prod_{h \in \psi} \tP_{\text{DEP},h}.
\end{align}

\subsection{Problem Formulation}

For a route $\psi =(h_1,\dots,h_{N_{\psi}})$ from a source to a destination, the single-hop link having the lowest throughput becomes the bottleneck of the entire route $\psi$.
Thus, we write the end-to-end throughput of route $\psi$ as
\begin{align} \label{eq:Rate_route}
	U(\psi) =  &\underset{ h \in \psi }{\min} ~~ U_h,
\end{align}
where $U_h$ is the throughput of a single-hop link $h$, which is expressed as 
\begin{align} \label{eq:Rate_link}
	U_h = \Omega^{(m_h)} \log_2\Bigg( 1 + \frac{P \big| g_{T_h,R_h}^{(m_h)}\big|^2 }{ \Omega^{(m_h)}N_0 } \Bigg).
\end{align}

Then, we can formulate the end-to-end DEP maximization problem with a constraint on end-to-end throughput requirement as
\begin{align} \label{eq:DEP_problem}
	&\underset{ \psi \in \Psi,  \left\{m_h\right\}_{h\in\psi }    }{\max} ~~ \tP_{\text{DEP}}(\psi)  \\  \label{eq:Thput_constraint}
	&\qquad\quad \text{s.t.} ~~~  U(\psi) \geq U_{\text{target}}, 
\end{align} 
where $U_{\text{target}}$ specifies the minimum required end-to-end throughput at the destination.
In this letter, we focus on the development of an algorithm that solves the problem in~\eqref{eq:DEP_problem} in a distributed fashion.
For comparison and validation purposes, a centralized method where a central node finds the optimal route and modalities based on the global channel information collected from all nodes will be discussed in Section~\ref{subsec:centralized}.

\section{Proposed Q-Covert Routing for HetNets}

In this section, we introduce a novel distributed Q-covert routing algorithm for the DEP maximization problem in~\eqref{eq:DEP_problem}. 
In the Q-covert routing strategy, each node chooses its next hop and the best modality for that hop using its estimate of the end-to-end DEP, which is calculated based on local feedback from neighboring nodes.

\subsection{Definitions of State, Action, and Cost} \label{subsec:Define}

Let us denote $\Phi$ as the set of all legitimate nodes, which are a source, a destination, and multiple other nodes from which relay
nodes can be chosen. The state space of a transmitter $T \in \Phi$ is defined as the set of all possible destination nodes. 
Hence, for transmitter $T$, the state space $\cS_{T}$ is given by 
\begin{align} \label{eq:State_space}
	\cS_{T} = \Phi \setminus \{ T \}.
\end{align}

The goal of each transmitter is to select a neighboring node and choose the optimal communication modality to transmit to the selected neighboring node. 
We represent $\tilde{\cA}_T$ as the set of all possible actions at transmitter $T$.
Then, $\tilde{\cA}_T$ is expressed as 
\begin{align*} 
	\tilde{\cA}_T = \{ a_{R_{T,1}}^{(1)},\cdots, a_{R_{T,1}}^{(M)}, \cdots, a_{R_{T,N_T}}^{(1)},\cdots, a_{R_{T,N_T}}^{(M)}   \},
\end{align*}
where $N_{T}$ stands for the number of neighboring nodes for $T$, and $a_{R_{T,i}}^{(m)}$ means the action that transmitter $T$ decides to send data to neighboring node $R_{T,i}$ with communication modality~$m$.

Let us denote $h(a)$ as the single-hop link with the transmitter/receiver pair and communication modality associated with an action~$a$.
Note that the constraint on the end-to-end throughput in~\eqref{eq:Thput_constraint} is fulfilled if the throughput of each hop meets the requirement, i.e., $U_h \geq U_{\text{target}}$ for $h \in \psi$.  
Therefore, the action space of transmitter $T$, which is composed of all possible throughput requirement satisfying actions at transmitter~$T$, is written as 
\begin{align} \label{eq:Action_space}
	\cA_T = \{ a \in  \tilde{\cA}_T :  U_{h(a)} \geq   U_{\text{target}} \}. 
\end{align}

Each transmitter selects an action from its own action space $\cA_T$ in~\eqref{eq:Action_space} to solve the end-to-end DEP maximization problem in~\eqref{eq:DEP_problem}. 
As in~\eqref{eq:DEP_route}, the end-to-end DEP of a route $\psi$ is the product of the DEPs of all hops in the route $\{ \tP_{\text{DEP},h} \}_{h\in\psi}$, and thus the end-to-end DEP maximization can be reformulated as 
\begin{align} \label{eq:DEP_route_problem_reformulation}
	\underset{ \psi \in \Psi   }{\max} ~~ \prod_{h \in \psi} \tP_{\text{DEP},h} &~\Leftrightarrow~ \underset{ \psi \in \Psi   }{\max} ~~ \sum_{h \in \psi} \ln \left( \tP_{\text{DEP},h} \right) \nonumber \\
     &~\Leftrightarrow~ \underset{ \psi \in \Psi   }{\min} ~~ \sum_{h \in \psi} \ln \left( \frac{1}{\tP_{\text{DEP},h}} \right).
\end{align} 
Based on this observation, we define the cost of an action $a \in \cA_T$ at transmitter $T$ as 
\begin{align} \label{eq:Cost}
	c_T(a) = \ln \left( \frac{1}{\tP_{\text{DEP},h(a)}} \right).
\end{align}

\subsection{Proposed Algorithm} \label{subsec:Proposed}

In Q-learning, a transmitter $T \in \Phi$ takes an action $a \in \cA_T$ at a state $s \in \cS_T$ leveraging its estimate on the end-to-end performance, which is called a Q-value. 
Then, the transmitter $T$ receives feedback information from the selected receiver and updates its Q-value based on the information.

Since the goal is to minimize the end-to-end performance metric in~\eqref{eq:DEP_route_problem_reformulation}, transmitter $T$ selects the action that has the lowest Q-value, i.e., $\hat{a} = \underset{ a \in \cA_T }{\arg\min} ~ Q_T(s,a)$ where $Q_T(s,a)$ accounts for the estimated DEP from transmitter $T$ to destination $s$ for action $a$.
Also, we employ the $\epsilon$-greedy method~\cite{Sutton:98} where transmitter $T$ decides its action based on the Q-value (exploitation) with probability $1-\epsilon$ and chooses any random action (exploration) with probability $\epsilon$ to explore new actions that have not been learned. 
Hence, transmitter $T$ selects an action as follows:
\begin{align} \label{eq:Txpower}
	\hat{a} = \left \{ \!\! \begin{array}{ll}  {\displaystyle \underset{ a \in \cA_T }{\arg\min} ~ Q_T(s,a) }   &\!\!\!\!\!\!\!\!\!\!\!\! \textrm{with probability} ~ 1 - \epsilon, \vspace{3mm}  \\   
					          {\displaystyle   \text{choose a random action from } \cA_T}  & \!\! \textrm{with probability}   ~ \epsilon.  \end{array}  \right.  
\end{align}

After taking an action $a_{R_{T,i}}^{(m)} \in \cA_T$, transmitter~$T$ receives feedback from the selected receiving node $R_{T,i}$. 
The feedback information consists of two values, i.e., the immediate cost and future cost. 
The immediate cost indicates the cost of the single-hop link between transmitter $T$ and $R_{T,i}$ using communication modality $m$, which is $c_T\big(a_{R_{T,i}}^{(m)}\big)$ in~\eqref{eq:Cost}.
On the other hand, the future cost specifies the expected DEP from $R_{T,i}$ to the destination, which is given by $\hat{c}_T\big(a_{R_{T,i}}^{(m)}\big) \triangleq \underset{ a \in \cA_{R_{T,i}}  }{\min}Q_{R_{T,i}}(s,a)$. 
Then, based on the received feedback information, transmitter $T$ updates its Q-value as follows:
\begin{align} \label{eq:Q-value_update}
	Q_T\Big(s,a_{R_{T,i}}^{(m)}\Big) & \leftarrow  (1-\alpha) Q_T\Big(s,a_{R_{T,i}}^{(m)}\Big)   \\
                    &~~~+ \alpha \left( c_T\Big(a_{R_{T,i}}^{(m)}\Big) + \gamma  \hat{c}_T\big(a_{R_{T,i}}^{(m)}\big) \right), \nonumber
\end{align}
where $\alpha \in [0,1]$ is the learning rate that determines how much newly received information will be adopted in the Q-value. 
Here, $\gamma \in [0,1]$ is the discount factor that controls the weight of the future cost.


The details of the proposed Q-covert routing algorithm for simultaneous route optimization and modality selection for each of the hops are provided in Algorithm~1 based on the defined state space in~\eqref{eq:State_space}, action space in~\eqref{eq:Action_space}, cost in~\eqref{eq:Cost}, and Q-value in~\eqref{eq:Q-value_update}.
For a given source/destination pair, the source first selects an action based on its Q-value.
Then, the chosen receiver feeds back information to the source and takes its own action to decide its next hop and modality.
This step is repeated until data arrives at the destination.
This process is called an \textit{episode} and is repeated $n_{\text{episode}}$ times where the Q-values are updated in every episode.
Lastly, a route and the modalities at all hops on the route are identified utilizing the obtained Q-values.

\vspace{-6mm}
\bea 
\begin{array}{l}
\vspace{-1mm}
\text{-------------------------------------------------------------------------} \\   \vspace{-1mm}
\text{\textbf{Algorithm~1:} Q-covert Routing Algorithm for HetNets} \\ \vspace{-1mm}
\text{-------------------------------------------------------------------------} \\  
\text{Define the source and destination as $S$ and $D$, respectively.}\\  
\text{Define $\cS_T$ in~\eqref{eq:State_space} and $\cA_T$ in~\eqref{eq:Action_space}, $\forall T \in \Phi$.}\\  
\text{Initialize Q-values as $Q_T(s,a) = 0, \forall T \in \Phi, \forall s$, and $\forall a$. }   \\ 

\text{\textbf{for} $n:=1$ to $n_{\text{episode}}$ \textbf{do} }   \\ 
\text{~~Set $\mu = 1$ and $T = S$.}   \\  
\text{~~\textbf{while} $\mu = 1$}   \\ 
\text{~~ ~~Generate a random number $r$ from a uniform }\\ 
\text{~~ ~~distribution with the interval $[0, 1]$. }\\ 
\text{~~ ~~\textbf{if} $r \leq \epsilon$ \textbf{then}}   \\ 
\text{~~ ~~ ~~Select an action $a_{R_{T,i}}^{(m)}$ randomly from $\cA_T$.}   \\ 
\text{~~ ~~\textbf{else}}   \\ 
\text{~~ ~~ ~~Select $a_{R_{T,i}}^{(m)}$ according to } \underset{ a_{R_{T,i}}^{(m)} \in \cA_T }{\arg\min} ~ Q_T(s,a). \vspace{-2mm} \\  
\text{~~ ~~\textbf{end if} }   \\ 
\text{~~ ~~Execute the action $a_{R_{T,i}}^{(m)}$ and receive the feedback } \\  
\text{~~ ~~$c_T\big(a_{R_{T,i}}^{(m)}\big)$ and $\hat{c}_T\big(a_{R_{T,i}}^{(m)}\big)$ from the receiver $R_{T,i}$.} \\  
\text{~~ ~~Update $Q_T\Big(s,a_{R_{T,i}}^{(m)}\Big)$ according to~\eqref{eq:Q-value_update}.}\\ 
\text{~~ ~~\textbf{if} $R_{T,i} = D$ \textbf{then}}   \\ 
\text{~~ ~~ ~~$\mu = 0$}   \\ 
\text{~~ ~~\textbf{else}}   \\ 
\text{~~ ~~ ~~Set $T=R_{T,i}$} \\  
\text{~~ ~~\textbf{end if} }   \\ 
\text{~~\textbf{end while} }   \\  
\text{\textbf{end for}}   \\  


\text{An end-to-end route and modalities for all single-hop}   \\  
\text{links are identified based on the obtained Q-values. }   \\

\text{-------------------------------------------------------------------------} \\ 
\end{array} \qquad \qquad \qquad \qquad \qquad \qquad \qquad\nonumber
\nonumber \eea

\subsection{Centralized Approach} \label{subsec:centralized}

Now, let us introduce the centralized method~\cite{Kong:24} for the problem in~\eqref{eq:DEP_problem}. 
First, the HetNet is modeled as a directed graph where the edge from a transmitter to a receiver exists if the throughput of the communication between them is not less than the target throughput $U_{\text{target}}$.
Following~\eqref{eq:DEP_route_problem_reformulation}, the end-to-end DEP maximization is equivalent to the minimization of $\sum_{h \in \psi} \ln \left( \frac{1}{\tP_{\text{DEP},h}} \right)$.
Thus, the problem in~\eqref{eq:DEP_problem} can be reformulated to the shortest path problem when the edge weight for link $h$ is defined by $\ln \left( \frac{1}{\tP_{\text{DEP},h}} \right)$.
Lastly, the optimal solution for the formulated shortest path problem is obtained by exploiting Dijkstra's algorithm~\cite{Medhi:17}. 
Note that the end-to-end DEP of the centralized technique can be considered as an upper bound of that of any decentralized routing scheme.

\section{Numerical Results} \label{sec:simul}

\begin{figure}[t]
\begin{center} \hspace*{-0.45cm}
\includegraphics[width=0.4\textwidth]{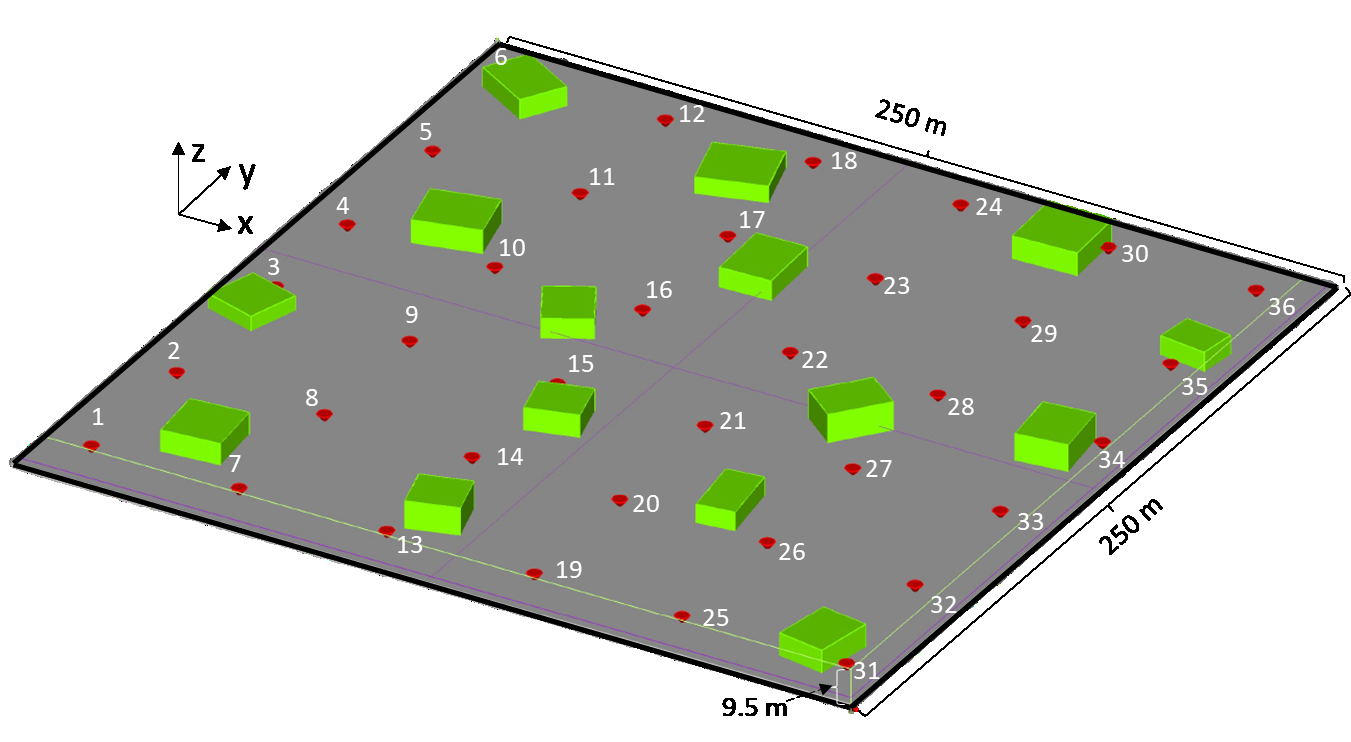}
\end{center} \vspace{-3.0mm}
\caption{A 3D simulation environment with 36 legitimate nodes.}  
\label{figure_Simualted_Scene}
\end{figure}

In this section, we provide numerical simulation results to show the efficiency of the developed Q-covert routing strategy.
As illustrated in Fig.~\ref{figure_Simualted_Scene}, we consider a 3D simulation environment (250$\times$250$\times$9.5) m$^3$  with multiple concrete buildings (green cuboids) and 36 legitimate nodes (red cones) having vertically polarized short dipole antennas at a height of 3~m above ground.
For this environment, channel data for three communication modalities $M_1$, $M_2$, and $M_3$ whose center frequencies are respectively 400~MHz, 900~MHz, and 2.4~GHz are obtained based on ray-tracing approaches with a commercial software~\cite{EMCUBE}.
Here, unless otherwise stated, we set $L=100$, $P=10$~dBm, $N_0=-80$~dBm, $\Omega^{(m)}=4$~MHz, $U_{\text{target}} = 0.5$~Mbps, $n_{\text{episode}}=300$, $\alpha = 0.3$, $\gamma = 0.9$, and $\epsilon = 0.1$. 
Also, the source and destination are nodes~1 and~36, respectively, and a subset of the other nodes are relay nodes. 

We compare the proposed algorithm with three routing methods. 
The first approach is \textit{the centralized scheme} in Section~\ref{subsec:centralized} which provides an upper bound of the maximum achievable end-to-end DEP with any centralized or decentralized routing approach.
Secondly, in \textit{the closest to the destination strategy}, each transmitting node selects the neighbor closest to the destination and chooses the modality exhibiting the highest DEP for the selected neighbor.
Lastly, \textit{the best direction to the destination} method indicates the case where each transmitter chooses the neighbor having the smallest angle offset between the transmitter-to-neighbor direction and the transmitter-to-destination direction, and selects the modality having the largest DEP for the chosen neighbor.
In the last two benchmark schemes, each transmitter selects only neighbors that are more than 50~m away from Willie to reduce the probability of being detected by Willie.

\begin{figure}[t]
\begin{center} \hspace*{-0.45cm}
\includegraphics[width=0.38\textwidth]{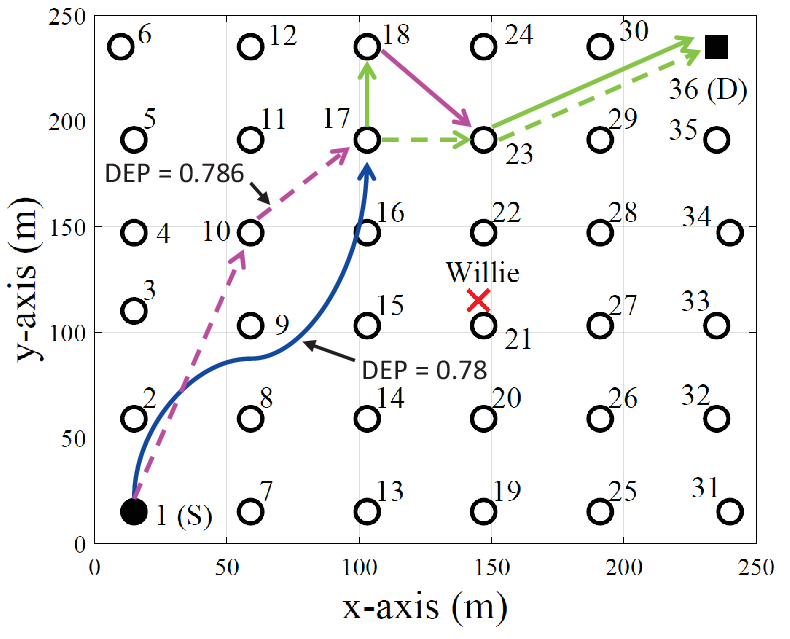}
\end{center} \vspace{-4.0mm}
\caption{Optimizes routes and selected modalities. The hops with modalities $M_1$, $M_2$, and $M_3$ are respectively highlighted in blue, magenta, and green. The routes with solid lines and dashed lines are from the proposed technique and the centralized method, respectively.}  
\label{figure_route}
\end{figure}


Fig.~\ref{figure_route} demonstrates the optimized routes and modalities from the proposed Q-covert routing algorithm and the centralized method when Willie is located at $(145,115)$.
It is seen that, in the proposed scheme, each transmitting node judiciously chooses the best modality based on its local channel characteristic. 
Another observation is that the optimized routes for both approaches tend to avoid using nodes that are close to Willie to increase the end-to-end DEP. 
In addition, the proposed routing technique provides the end-to-end DEP 0.78, which is very close to the end-to-end DEP of the centralized approach which is 0.786.

\begin{figure} 
    \centering \hspace*{-0.4cm} 
    \subfigure[Target throughput $U_{\text{target}}=0.5$~Mbps]{\label{figure_2_1}\includegraphics[width=0.55\textwidth]{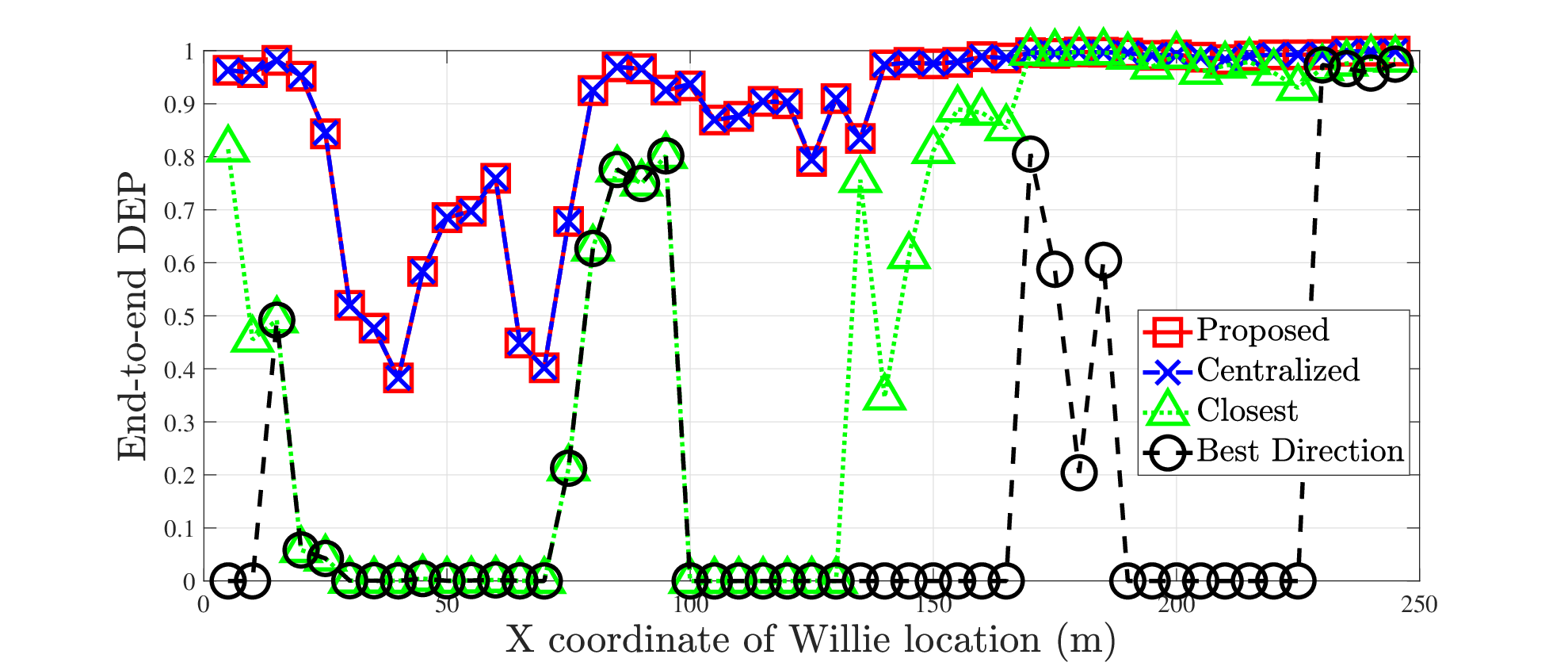}}  \hspace*{-0.4cm}   
    \subfigure[Target throughput $U_{\text{target}}=1.5$~Mbps]{\label{figure_2_1}\includegraphics[width=0.55\textwidth]{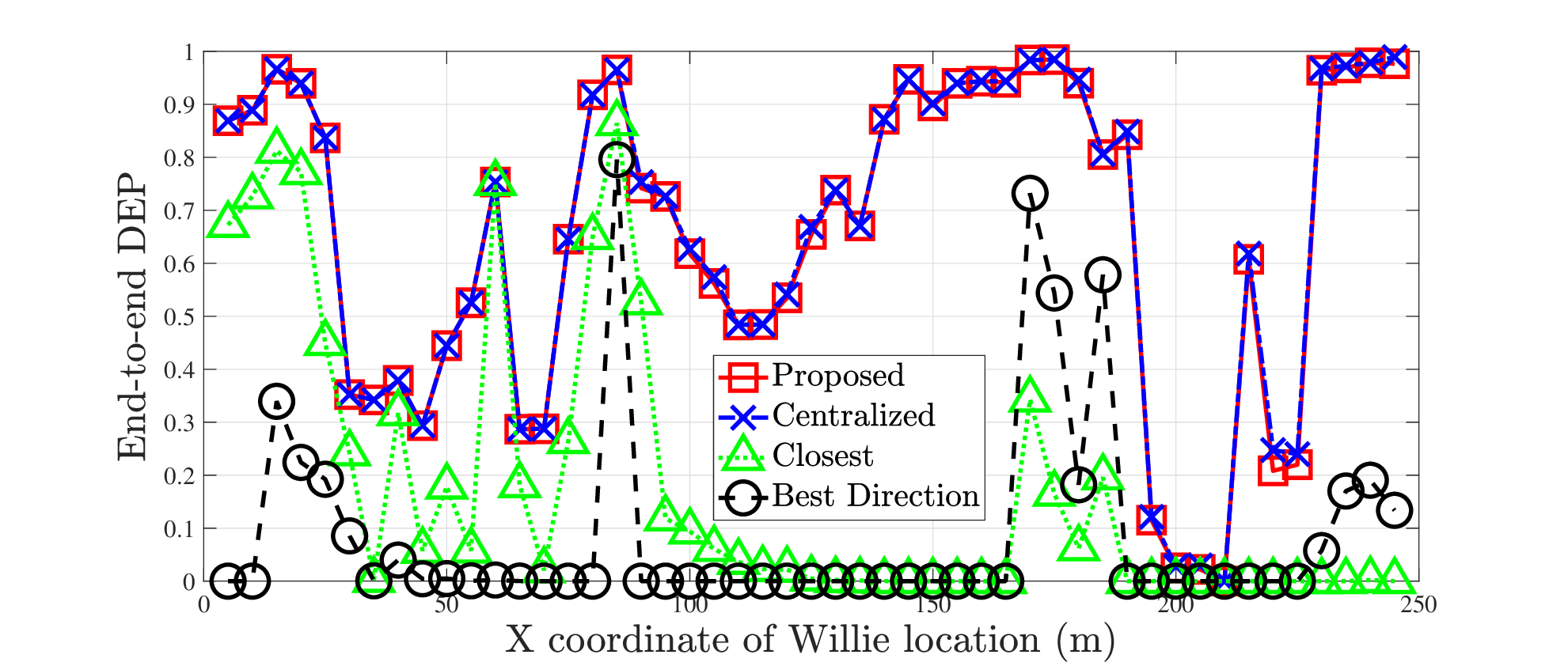}}
    \vspace{-3.0mm}
    \caption{The end-to-end DEP performance as a function of the X coordinate of Willie's location.}  
    \label{figure_DEP}
\end{figure}

In Fig.~\ref{figure_DEP}, we present the end-to-end DEP performance as a function of the X coordinate of Willie $L_{W,X}$ for two values of the target throughput $U_{\text{target}}$ where Willie is located at $(L_{W,X},125)$. First, it is shown that the performance varies dramatically with the position of Willie. 
Also, we can observe that the proposed Q-covert routing strategy exhibits almost identical performance to the centralized method and significantly outperforms the two benchmark schemes.
Lastly, as expected from the problem in~\eqref{eq:DEP_problem}, the end-to-end DEP performance is highly dependent on $U_{\text{target}}$, and it becomes smaller when $U_{\text{target}}$ gets larger.

\begin{figure}[t]
\begin{center} \hspace*{-0.5cm}
\includegraphics[width=0.55\textwidth]{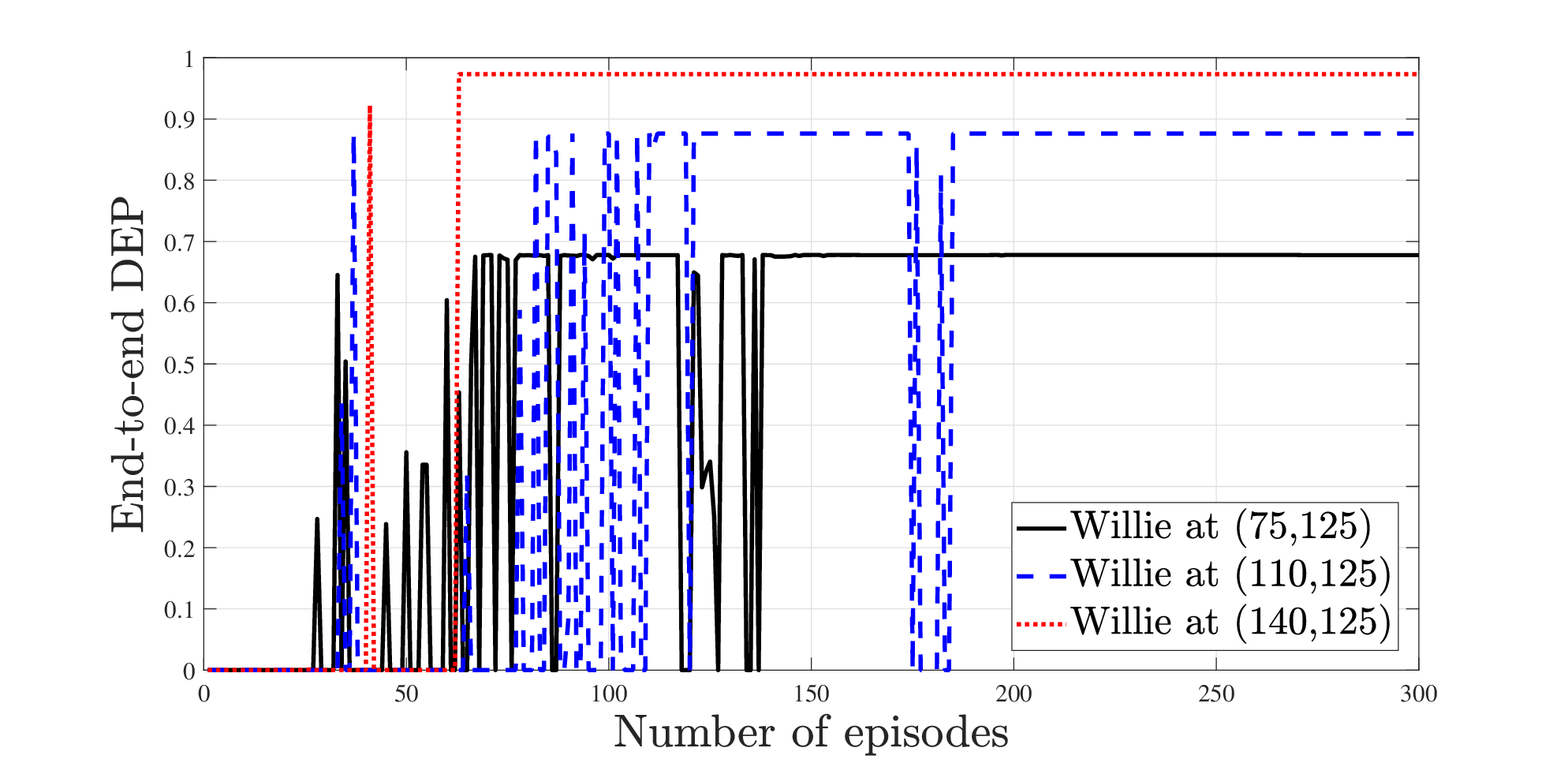}
\end{center} \vspace{-3.0mm}
\caption{The end-to-end DEP performance of the proposed Q-covert routing algorithm as a function of the number of episodes $n_{\text{episode}}$.  } 
\label{figure_converge}
\end{figure}


In Fig.~\ref{figure_converge}, we investigate the convergence behavior of the proposed Q-covert routing algorithm for various locations of Willie.
When the number of episodes $n_{\text{episode}}$ is small, since each node estimates the end-to-end DEP only based on a small number of feedback, the estimation may not be accurate, and this leads to the degraded DEP. 
When the number of episodes increases, the performance of the Q-covert routing technique improves and it converges to a constant value. Fig.~\ref{figure_converge} shows that the proposed strategy converges for all cases when $n_{\text{episode}} = 200$.

\section{Conclusion}

In this letter, we have examined the joint route and modality selection problem for covert communications in HetNets. 
We have developed a novel decentralized Q-covert routing algorithm that identifies a route from a source to a destination and selects a modality for each of the transmitting nodes in the route with the goal of maximizing the end-to-end DEP while fulfilling a requirement on the end-to-end throughput.  
We have validated the efficacy of the proposed technique through extensive numerical simulations.

\bibliographystyle{ieeetr}
\bibliography{bibliography.bib}
\end{document}

%% file: macro_by_Justin.tex
\definecolor{orange}{rgb}{1.0, 0.5, 0.0}

\newcommand{\CC}{\mathbb C} 
\newcommand{\PP}{\mathbb P} 

\newcommand{\cA}{{\mathcal A}}

\newcommand{\cC}{{\mathcal C}}
\newcommand{\cD}{{\mathcal D}}

\newcommand{\cH}{{\mathcal H}}

\newcommand{\cN}{{\mathcal N}}

\newcommand{\cS}{{\mathcal S}}

\newcommand{\tP}{{\texttt P}}

\renewcommand{\d}{\operatorname{d\!}{}} 

\newcommand{\be}{\begin{equation}}
\newcommand{\ee}{\end{equation}}
\newcommand{\bea}{\begin{eqnarray}}
\newcommand{\eea}{\end{eqnarray}}
\newcommand{\bitem}{\begin{itemize}}
\newcommand{\eitem}{\end{itemize}}

\newcommand{\bdp}{\begin{displaymath}}
\newcommand{\edp}{\end{displaymath}}